# High-efficiency light-wave control with all-dielectric optical Huygens' metasurfaces

*Manuel Decker[1], Isabelle Staude[1,]\*, Matthias Falkner[2], Jason Dominguez[3], Dragomir N. Neshev[1], Igal Brener[3], Thomas Pertsch[2], and Yuri S. Kivshar[1]*

[1]Nonlinear Physics Centre, Research School of Physics and Engineering, The Australian National University, Canberra ACT 0200, Australia

[2]Institute of Applied Physics, Abbe Center of Photonics, Friedrich-Schiller-Universität Jena, 07743 Jena, Germany

[3]Center for Integrated Nanotechnologies, Sandia National Laboratories, Albuquerque, New Mexico 87185, USA

\*E-mail: isabelle.staude@anu.edu.au

**Optical metasurfaces have developed as a breakthrough concept for advanced wave-front engineering enabled by subwavelength resonant nanostructures. However, reflection and/or absorption losses as well as low polarisation-conversion efficiencies pose a fundamental obstacle for achieving high transmission efficiencies that are required for practical applications. Here we demonstrate, for the first time to our knowledge, highly efficient all-dielectric metasurfaces for near-infrared frequencies using arrays of silicon nanodisks as meta-atoms. We employ the main features of Huygens' sources, namely spectrally overlapping electric and magnetic dipole resonances of equal strength, to demonstrate Huygens' metasurfaces with a full transmission-phase coverage of 360 degrees and near-unity transmission, and we confirm experimentally full phase coverage combined with high efficiency in transmission. Based on these key properties, we show that all-dielectric Huygens' metasurfaces could become a new paradigm for flat optical devices, including beam-steering, beam-shaping, and focusing, as well as holography and dispersion control.**

More than 300 years ago Huygens proposed that every point on a wave front acts a secondary source of outgoing waves [1]. Notably, despite its simplicity, Huygens' Principle already allows for the derivation of the fundamental laws of diffraction and reflection and intuitively explains how components like conventional lenses, which are based on the accumulation of (position-dependent) phase delay during wave propagation, can be used to shape an emerging wave front in a desired way. A more rigorous formulation of the Huygens' principle developed by Love more than 200 years later [2] reveals that in order to achieve purely forward-propagating elementary waves, as required by the Huygens' principle, each individual elementary source should be described as an electrically small antenna that radiates the far-fields of crossed electric and magnetic dipoles (Huygens sources) [2-4]. While these elementary sources were originally introduced as fictitious entities, an actual physical implementation of Huygens' sources can be achieved using polarizable subwavelength particles that sustain both electric and magnetic dipolar resonances [3-6]. Arranging many of such particles in a plane furthermore allows for creating effective Huygens' surfaces - reflectionless sheets that can be used to manipulate electromagnetic waves by controlling the resonant properties of the particles as a function of position [5,6].

Only recently, experimental demonstrations of metallic Huygens surfaces at microwave frequencies [5], where the dissipative losses of metals are negligible, have been presented, and design concepts that allow bringing plasmonic Huygens' surfaces to the mid-infrared spectral range have also been developed [6]. However, transferring this powerful concept to near-infrared and visible wavelengths remains unsolved. This challenge originates from the weak magnetic response of natural materials at optical frequencies. Even for plasmonic nanoparticles, despite their capability to provide a magnetic permeability that significantly deviates from unity [7,8], the critical combination of spectrally overlapping pure magnetic and electric dipole resonances of equal strength, as it is required for resonant Huygens sources, has not been realized yet. Furthermore, downscaling the suggested structures [5,6] to near-infrared or even visible operation wavelengths is further hindered by the significant increase of dissipative losses of plasmonic structures at optical frequencies. Overcoming these limitations is essential for realizing Huygens' surfaces at near-infrared operation wavelengths.

We here present and demonstrate a new route for successfully implementing Huygens' sources at optical frequencies that utilizes the strong resonant response from high-permittivity all-dielectric nanoparticles in the visible and near-infrared spectral range [9-18]. Such nanoparticles support both electric and magnetic dipolar Mie-type modes while, at the same time, exhibiting very low intrinsic losses [19-20]. The key importance of our solution is that the nanoparticles' magnetic dipole resonances can be pure and of comparable strength as their electric counterparts, making them ideal candidates for Huygens' sources. This allows us to realize effective all-dielectric Huygens' metasurfaces, *i.e.*, subwavelength lattices of Huygens' sources or meta-atoms, that stand out by the absence of dissipative losses and their ability to eliminate unwanted reflections. At the same time they provide full $2\pi$ coverage for the phase shift imposed onto a transmitted light wave without relying on cross-polarization schemes as most of the plasmonic counterparts [21-27]. Consequently, Huygens' metasurfaces in principle allow for wave-front manipulation with ideally 100% transmission efficiency and can also provide almost arbitrary spatial distributions of phase discontinuities depending on the design of the meta-atoms at each position. The latter aspect is of particular interest for wave-front manipulation applications. Indeed, it has been shown recently that tailoring the spatial distribution of phase discontinuities with optical metasurfaces allows for the realization of a flurry of flat optical devices [21-31]. However, the big obstacle in rendering these (mostly plasmonic) devices practical for real-world applications is their low efficiency, which stems from reflections, dissipative losses, and/or low polarization-conversion yield. All-dielectric optical Huygens' metasurfaces eliminate these restrictions and, hence, can be used for high-efficiency wave-front manipulation at near-infrared frequencies. To demonstrate this concept experimentally we employ silicon nanodisks as meta-atoms, which we arrange in a two-dimensional subwavelength array. By direct phase-resolved transmittance measurements we experimentally demonstrate the characteristic properties of a Huygens' metasurface, *i.e.*, a complete phase coverage from 0 to $2\pi$ for nanodisks with spectrally overlapping resonances accompanied by very high transmittance efficiency. Our experimental results are supported by numerical calculations, indicating that near-unity efficiency can be achieved for a corresponding metasurface embedded in an appropriate dielectric environment.

1. **Analytical model of a resonant Huygens' metasurface**

In order to gain a fundamental understanding of the optical response of a resonant Huygens' metasurface, we first discuss the full complex response of an idealized sub-wavelength array of loss-less all-dielectric nanodisk meta-atoms with electric and magnetic dipole resonances of equal strength and width. Our analytical model is based on a coupled discrete dipole approach [32] where each individual silicon nanodisk is represented by an ideal Huygens source, *i.e.*, a pair of electric and magnetic dipoles oriented in *x* and *y* direction with specific polarizabilities that follow a Lorentzian frequency dependence with resonance positions $\omega_{e,0}$ and $\omega_{m,0}$ and damping parameters $\gamma_e$ and $\gamma_m$, respectively. The two dipoles are then arranged in an infinite sub-wavelength 2D-square lattice forming the silicon-nanodisk Huygens' metasurface (Fig. 1 a). Using this approach we calculate an analytical expression for the (field-) transmission and reflection coefficients as a function of frequency $\omega$ (see Methods), yielding

$$t = 1 + \frac{2i \cdot \gamma_e \cdot \omega}{\omega_{e,0}^2 - \omega^2 - 2i\gamma_e\omega} + \frac{2i \cdot \gamma_m \cdot \omega}{\omega_{m,0}^2 - \omega^2 - 2i\gamma_m\omega} \qquad (1)$$

and

$$r = \frac{2i \cdot \gamma_e \cdot \omega}{\omega_{e,0}^2 - \omega^2 - 2i\gamma_e\omega} - \frac{2i \cdot \gamma_m \cdot \omega}{\omega_{m,0}^2 - \omega^2 - 2i\gamma_m\omega}, \qquad (2)$$

respectively. Importantly, the last terms in Eqs. 1 and 2 stem from the contribution of the magnetic dipoles in the lattice whereas the electric contributions are associated with the preceding terms. Evidently, the sign of the electric field component of the magnetic-dipole mode radiated in forward *versus* backward direction is different while the sign of the electric-field component of the electric-dipole resonance is the same for both directions. This aspect is, indeed, a known feature discriminating magnetic and electric dipole radiation and can be utilized to create, *e.g.*, an optical magnetic mirror [33,34] in reflection. The total response in transmittance consists of the interference of the collective magneto-electric response of the metasurface and the incident plane wave.

Figure 1 b,c shows the full (amplitude and phase) spectral response (green) obtained by our analytical approach for an ideal loss-less metasurface with spectrally separate electric ($\lambda_{el}$ = 1,400 nm) and magnetic ($\lambda_{mag}$ = 1,200 nm) resonances of equal strength and width. Interference with the incident plane wave results in the expected two transmittance dips shown as green line in Fig. 1 b. Correspondingly, the phase response of the transmitted field amplitude shows resonant behaviour at the corresponding resonance wavelengths with a maximum phase change in resonance of $\pi$ for both resonances. Figure 1 d further visualizes the resonance signature in terms of an electric-field vector diagram for the non-overlap case at a selected wavelength ($\lambda$ = 1,252 nm).

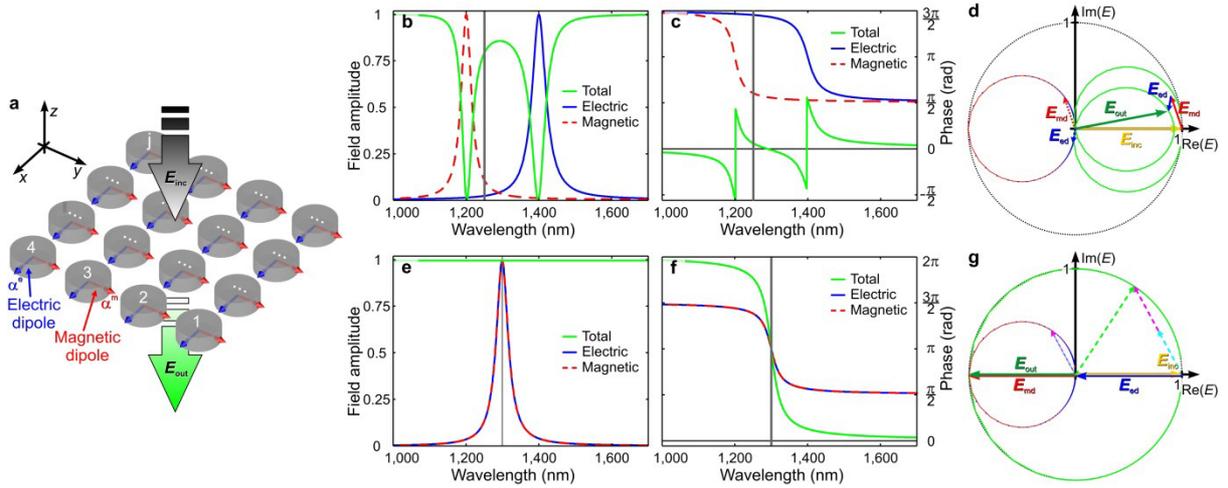

**Figure 1.** (a) Schematic of an (infinite) array of nanodisks represented as electric and magnetic dipoles with the polarizabilities $\alpha^e$ and $\alpha^m$ for x-polarized incident light. (b) Field amplitude of the transmitted electric field (green) and the contributions of the electric (blue) and magnetic (red) resonances of the metasurface for the case of non-overlapping resonances. (c) Phase spectrum of the transmitted electric field (green) and the electric (blue) and magnetic (red) response. (d) Vector diagram depicting the decomposition of the transmitted electric-field vector ($E_{out}$, green) into the contributions from the electric ($E_{ed}$, blue) and magnetic ($E_{md}$, red) resonances and the incident plane wave ($E_{inc}$, yellow) at $\lambda = 1,252$ nm (vertical gray line in b,c). The transmitted field is depicted as green line. (e-g) Corresponding diagrams for the case of spectrally overlapping electric and magnetic resonances: (e) Field amplitudes, (f) phase spectrum, and (g) vector diagram at $\lambda_{el} = \lambda_{mag} = 1,300$ nm. The electric and magnetic field components for a second exemplary wavelength are also plotted (cyan and magenta).

Figures 1 e-f show the corresponding diagrams for the case of spectrally overlapping electric and magnetic resonances of equal strength and width. In sharp contrast to the case of separate resonances, we observe a flat line with a transmittance of unity in the transmitted field amplitude for this ideal case. More importantly, the phase response of the total electric field changes drastically as evident from Fig. 1 f and undergoes a phase change of $2\pi$ within 200 nm around the resonance wavelength – this is double the phase shift that can be achieved by a single (electric or magnetic) resonance. Notably, although the resonant behaviour observed in the phase of the transmitted light is not reflected in its amplitude, our results are fully consistent with causality as expressed by the Kramers-Kronig relations. In order to visualize how the contributions of the electric and the magnetic resonances in combination with the incident plane wave lead to unity resonant transmission we again plotted the vector diagram of the electric field components in Fig. 1 g for the resonant case at $\lambda = 1,300$ nm. Evidently, the electric field component of the incident plane wave is eliminated by destructive interference with one of the two dipole resonances of the silicon-nanodisk metasurface while the second dipole resonance provides unity field amplitude that is out-of-phase with the incident plane wave (red and blue arrows). The sum of the electric field components of the plane wave and of the electric and magnetic resonances lie on an $|\mathbf{E}_{\text{out}}| = 1$ circle independent of wavelength, which is equivalent to the flat $T = 1$ line in transmission shown in Fig. 1 e. This unity transmission efficiency for zero absorption can also be described in terms of resonant impedance matching in metasurfaces with effective electric and magnetic polarizabilities of equal strength. Importantly, this behaviour cannot be observed for any single dipolar resonance but is characteristic for the interference of electric and magnetic dipole resonances of equal strength. As such our results are a direct manifestation of the magnetic-dipole radiation characteristics of the magnetic resonance of the silicon nanodisks.

## 2. Experimental results

In order to experimentally realize low-loss Huygens' metasurfaces for near-infrared frequencies we choose silicon nanodisks as meta-atoms since they allow for tailoring the spectral positions of their electric and magnetic dipole-type modes with respect to each other *via* controlling their aspect ratio [14]. Furthermore, silicon is an attractive material choice owing to its extremely low dissipative losses above 1.1 μm wavelength and its prevalence in modern technology.

We use standard silicon-on-insulator technology (see Methods) to fabricate subwavelength lattices of silicon nanodisks with different disk radii $r_d$ on a 2-μm-thick silicon-oxide layer and investigate the three cases of (1) mode overlap ($r_d \approx 242$ nm), (2) separate electric and magnetic resonances at the short-wavelength side of the overlap ($r_d \approx 198$ nm) and (3) separate resonances at the long-wavelength side of the overlap ($r_d \approx 310$ nm). The lattice constant $a$ of the samples (1) and (2) is 666 nm and 913 nm for sample (3). A sketch of the sample geometry and scanning-electron micrographs of a typical sample with a disk radius of 242 nm are shown in Fig. 2.

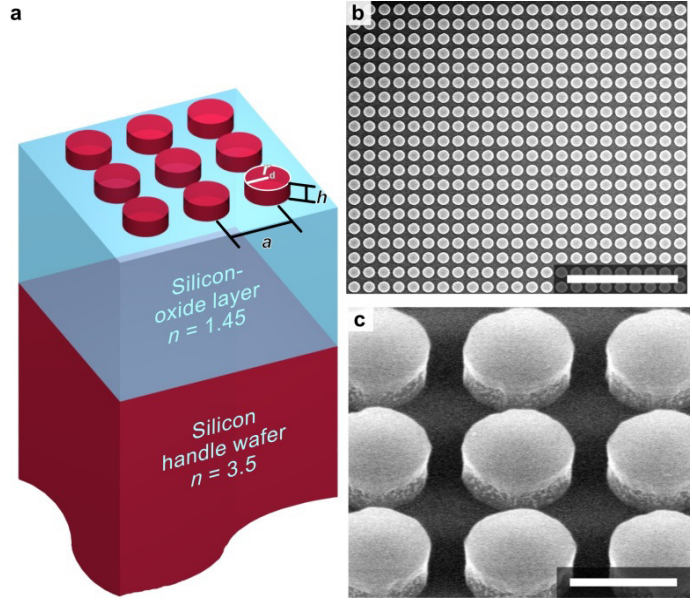

**Figure 2.** (a) Schematic of the silicon nanodisk metasurface on top of a 2μm-thick silicon oxide layer and a bulk silicon substrate. (b,c) Top-view and side-view electron micrographs of a silicon nanodisk sample with a disk radius of $r_d \approx 242$ nm, disk height of $h = 220$ nm, and a lattice constant of $a \approx 666$ nm. The scale bar in (b) is 5 μm and in (c) 500 nm.

We then measure the samples' intensity transmittance spectra using a conventional white-light spectroscopy setup and the spectrally-resolved transmittance phase with an interferometric white-light spectroscopy setup [35], respectively (see Methods). Our experimental results are summarized in Fig. 3 a-c. Note that the 2-μm-thick buried silicon-oxide layer that separates the silicon nanodisks from the silicon handle wafer (see Fig. 2 a) gives rise to Fabry-Perot oscillations in the transmittance spectra (insets of Fig. 3 a-c).

Figure 3 b experimentally confirms the predicted strong phase response covering the full range from 0π to 2π within a wavelength range of ~200 nm in the measured transmittance-phase response. In this case the electric and magnetic resonances of the silicon-nanodisk metasurface are close to the spectral overlap condition for a nanodisk radius of 242 nm. This is, indeed, the unambiguous fingerprint of a Huygens' surface. For the non-overlap cases ($r_d = 198$ nm and $r_d = 310$ nm) we find two typical single-resonance signatures in the transmittance-phase spectra (Fig. 3 a,c), each featuring a maximum phase change of π. Correspondingly, the measured intensity-transmittance spectra show pronounced spectrally separate resonances which can be identified as the electric and magnetic dipolar modes of the nanodisk metasurface [14], respectively. For the case of the smallest disk radius ($r_d = 198$ nm) the electric resonance occurs at $\lambda_{el,198} = 1{,}105$ nm and the magnetic resonance at $\lambda_{mag,198} = 1{,}166$ nm (Fig. 3 a). For the case close to the mode overlap ($\lambda_{el,242} \approx \lambda_{mag,242} \approx 1{,}230$ nm) no clear resonances can be identified (Fig. 3 b). Instead, the curve is dominated by Fabry-Perot oscillations of the layered wafer structure. As expected, for the largest disk radius of 310 nm the two resonances reappear (Fig. 3 c), the electric resonance now being the fundamental (low-energy) mode of the metasurface at $\lambda_{el,310} = 1{,}515$ nm while the magnetic resonance occurring at $\lambda_{mag,310} = 1{,}440$ nm becomes the high-energy mode. Our experimental results are in excellent agreement with our analytical predictions shown in Fig. 1.

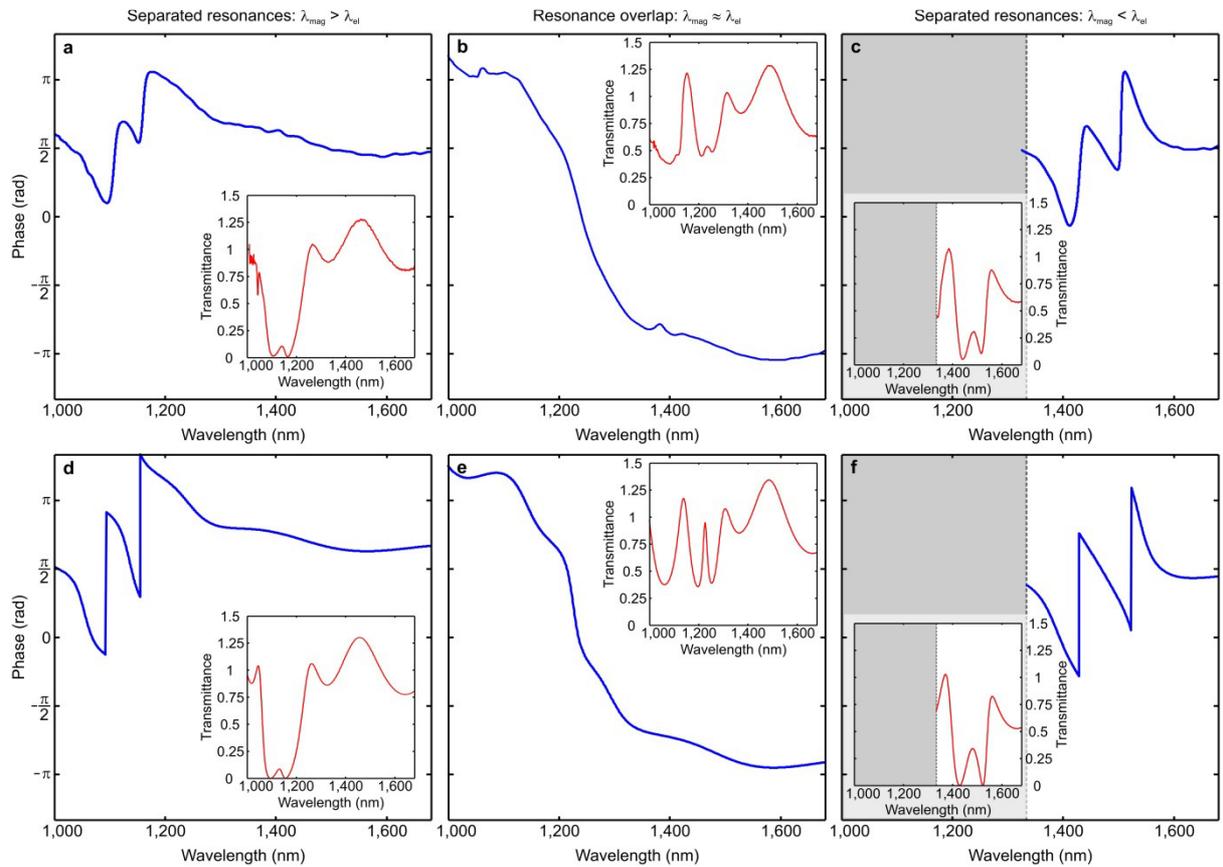

**Figure 3.** Measured transmittance-phase spectra (blue lines) of the nanodisk metasurfaces with (a) 198 nm, (b) 242 nm, and (c) 310 nm nanodisk radius. The corresponding transmittance intensity spectra $T/T_0$ are shown in the insets (red lines). The gray area in (c) depicts the part of the spectrum where diffraction into the lattice modes occurs due to the larger lattice constant chosen for this sample as compared to (a) and (b). (d-f) Corresponding numerical calculations for the three cases shown in (a-c). The transmittance-intensity spectra were referenced to the unstructured etched wafer next to the sample, resulting in nominal transmittance values larger than 1 due to its layered structure.

To further underpin our analytical and experimental findings, we additionally compare our measurements with numerical calculations using a finite element software package (see

Methods). These results are displayed in Fig. 3 d-f and show excellent agreement with the experimental spectra (Fig. 3 a-c) in all cases. Particularly, the numerical calculations again confirm the $2\pi$ phase change for spectrally overlapping modes that has been found in the experiment and the analytical model. Notably, near-unity transmittance, that is equivalent to resonant impedance matching of the metasurface to the dielectric environment, can be achieved for resonance overlap for an appropriate choice of the embedding medium. For demonstrating this aspect we remove the influence of the handle wafer in our numerical calculations – this can be performed experimentally, *e.g. via* a deep reactive ion etching process – and subsequently embed the Huygens' metasurface in a dielectric medium with a refractive index of $n = 1.66$. These results are shown in Fig. 4 a,b, showing the total transmission spectra of this system (colour coded) for a variation of the nanodisk radius from 120 nm to 320 nm at a constant nanodisk height of $h = 220$ nm. The lattice constant is set as $a = 1.38 \cdot 2 \cdot r_d$. For the incident light polarized in *x* direction and a disk radius of 306 nm, two spectrally separate resonances are clearly visible at $\lambda_{el,306} \approx 1650$ nm and $\lambda_{mag,306} \approx 1590$ nm wavelength, which can be identified as the electric and magnetic dipole-type modes, respectively. The condition of spectral overlap ($\lambda_{el} \approx \lambda_{mag}$) is met for $r_d \approx 242$ nm resulting in a nearly flat spectrum with a resonant transmittance value larger than 99% – naturally near-unity transmittance is equivalent to near-zero absorption and reflection. As expected, the $2\pi$ phase coverage is preserved. We note that these optimized conditions can be realized using, *e.g.*, transparent high-index polymers [36].

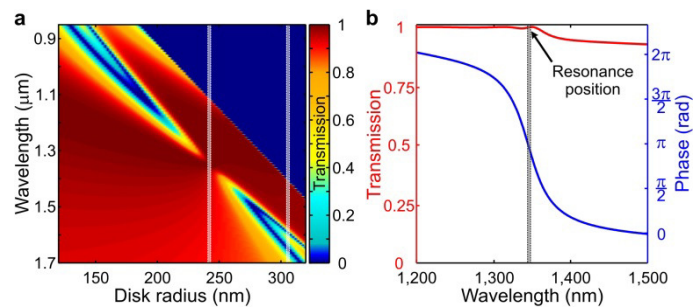

**Figure 4.** (a) Numerically calculated transmission spectra (colour coded) for a variation of the nanodisk radius $r_d$ from 120 nm to 320 nm at a constant nanodisk height of $h = 220$ nm for silicon-nanodisk metasurfaces embedded in a homogeneous medium with $n = 1.66$. The white-shaded lines indicate the spectra of the nanodisk radii for separate electric and magnetic resonances ($r_d = 306$ nm) and for the overlap case ($r_d = 242$ nm) also shown in (b). (b) Total transmission intensity (red solid line) and phase response (blue solid line) for spectral resonance overlap at $\lambda_{el,n=1.66} \approx \lambda_{mag,n=1.66} \approx 1,340$ nm featuring a transmission above $T_{min} \approx 0.995$ and full phase coverage of $2\pi$.

## 3. Wave-front engineering and dispersion control

In the following, we provide a glimpse of the technical potential that could be unlocked by all-dielectric optical Huygens' metasurfaces by presenting numerical and analytical model calculations for two different application examples:

First, for efficient wave-front manipulation and realization of devices like flat and lightweight beam deflectors (Fig. 5 a) and lenses, the capability to create a desired discrete phase gradient at a given operation wavelength is of paramount importance. For illustration of this concept, we choose the example of a linear phase gradient, as required for *e.g.* a beam deflector. We calculate spectrally-resolved relative phase delays for a systematic scaling of the nanodisks (Fig. 5 b) and detail the phase delays accessible at a wavelength of $\lambda \approx 1{,}358$ nm (Fig. 5 c). We then exemplarily use a 5-element, equidistant phase discretization and identify the corresponding nanodisk radii required for full phase coverage. Remarkably, the transmission efficiency for this case is larger than $T = 96\%$ over the entire $2\pi$ phase range (inset in Fig. 5 c).

The second example targets applications such as multi-photon microscopy and lithography, where the dispersion introduced by the focusing optics leads to a critical broadening of the pulse width of the exciting laser. Hence, prism, grating, or multiple-cycle chirped-mirror pulse compressors are required that are usually bulky and rely on precise (angular) alignment. We analytically show that an all-dielectric Huygens' metasurface (model parameters: $\lambda_{el} = \lambda_{mag} \approx 811$ nm and $\gamma_e = \gamma_m = 4$ THz) can be utilized as a thin, compact, and highly efficient pulse-compression device (Fig. 5 d) since it provides a strong group-delay dispersion of more than -2000 fs$^2$ (Fig. 5 e) in combination with unity transmittance. For our calculations, we concentrate on the practically important case of compressing a chirped 120-fs laser pulse at $\lambda = 800$ nm wavelength. Figure 5 f shows that a 120 fs chirped pulse – the chirp corresponds to propagation through 20 cm of glass, *e.g.*, in a typical two-photon microscope – is almost completely recompressed to the Fourier-limited initial pulse after transmission through 6 layers of our Huygens' metasurface. This shows that Huygens' metasurfaces can also provide remarkable dispersion-control that can be integrated in a linear beam path as an ultra-thin transmitting element.

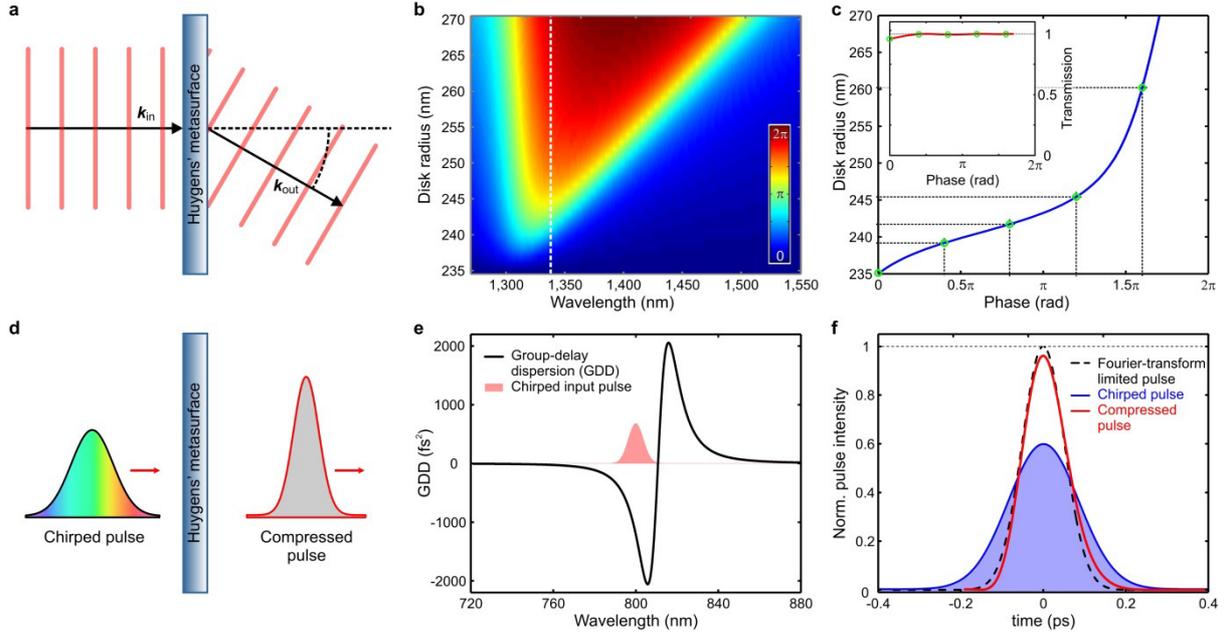

**Figure 5.** (a) Schematic of a Huygens' metasurface operating as a unity-efficiency beam deflector. (b) Calculated spectrally-resolved, relative phase delay (colour coded) of silicon-nanodisk Huygens' metasurfaces with different radii $r_d$ ($r_d$ = 235 nm to $r_d$ = 267 nm, constant aspect ratio $r_d/h$ = 1.21, embedding medium $n$ = 1.66) referenced to the phase delay for $r_d$ = 235 nm. (c) Range of phase delays accessible at a wavelength of $\lambda \approx 1{,}338$ nm (white-dashed line in b). A selection of nanodisk radii that provide a linear phase gradient, as required for, *e.g.* for beam deflection, with a 5-element discretization are plotted over the corresponding phase-delay values. The transmission values (red line) at these points (green circles) are shown in the inset. (d) Scheme showing the recompression of a chirped pulse after passing the Huygens' metasurface. (e) Group-delay dispersion (GDD) for an ideal Huygens' metasurface resonant at $\lambda_{el} = \lambda_{mag} \approx 811$ nm (black solid line) and the spectrum of the considered chirped laser pulse (red-shaded area). (f) Temporal profile of a 120-fs Fourier-limited pulse, of the chirped pulse, and of the recompressed pulse after propagation through 6 layers of the Huygens' metasurface.

## 4. Conclusions

In summary, we have experimentally demonstrated, for the first time to our knowledge, an all-dielectric Huygens' metasurface operating at near-infrared frequencies. Direct measurements of the transmittance phase reveal that the realized Huygens metasurface, which consists of a subwavelength array of silicon-nanodisk meta-atoms, provides full phase coverage from 0 to $2\pi$ while, at the same time, showing high transmittance. We further demonstrate that transmission efficiencies very close to 100% can be achieved for a realistic choice of the embedding dielectric medium. We analytically identify that the unique properties of our Huygens' metasurface have their foundation in the resonant overlap of electric and magnetic dipole resonances of the silicon-nanodisk resonators. Our results establish all-dielectric Huygens' metasurfaces as ideal candidates for near-unity-efficiency wave-front manipulation providing full phase coverage with negligible reflection and absorption losses. Therefore, they offer a practical route for the implementation of a multitude of transmitting optical devices, including flat optics, optical holography and dispersion control applications.

Importantly, our proposed Huygens' metasurfaces are fully compatible with standard industrial silicon technology and, due to the simple disk-shape of the meta-atom, also suitable for large-area fabrication techniques like interference lithography, nano-imprint lithography, and conventional optical lithography.

**Acknowledgements**

We acknowledge useful discussions with W. Liu, A. E. Miroshnichenko, M. Wegener, J. Fischer, D. Powell, V. Stoev, and A. Evlyukhin. This work was performed, in part, at the Center for Integrated Nanotechnologies, an Office of Science User Facility operated for the U.S. Department of Energy (DOE) Office of Science. Sandia National Laboratories is a multi-program laboratory managed and operated by Sandia Corporation, a wholly owned subsidiary of Lockheed Martin Corporation, for the U.S. Department of Energy's National Nuclear Security Administration under contract DE-AC04-94AL85000. The authors also acknowledge a support from the Australian Research Council, the Group of Eight: Australia -

**Methods**

**Fabrication**

For fabrication we have performed electron-beam lithography (EBL) on silicon-on-insulator wafers (SOITEC, 220 nm top silicon thickness, 2 μm buried oxide thickness, backside polished) using the negative-tone resist NEB-31A and HDMS as adhesion promoter. Development was performed by inserting the sample into MF-321 developer. The resulting resist pattern was used as an etch mask for a directive reactive-ion etching process. Residual resist was removed using oxygen plasma. The sample footprints are 2 mm x 2 mm.

**Transmittance measurements**

The intensity-transmittance spectra were collected by a home-built white-light spectroscopy setup connected to an optical spectrum analyzer. A tungsten halogen lamp was used as a broadband light source. The range of incident angles has been reduced to ±6 degrees by an aperture.

For transmittance-phase measurements we used a home-built interferometry setup described in more detail in Ref. [35]. In this setup a white-light beam from a supercontinuum source is divided into two parallel beams referred to as sample and reference beams by a polarizing beam displacer. The wafer with the silicon nanodisk fields was inserted into the beam path in such a way that the sample beam was passing through the respective field to be measured and the reference beam was passing through an unstructured area of the wafer next to the field. After transmission through the wafer the two beams were recombined using a second polarizing beam displacer and sent to a spectrometer.

**Analytical coupled-dipole model**

Each silicon-nanodisk meta-atoms is modelled as electric and magnetic dipoles oriented in $x$ and $y$ direction with specific polarizabilities $\alpha^e$ and $\alpha^m$ that are arranged in an infinite sub-wavelength 2D-square lattice (see Fig. 1 a). For $x$-polarized illumination with a plane wave propagating in $-z$ direction $\boldsymbol{E}(z,t) = exp(-ik_d z - i\omega t)$, where $k_d$ is the wavevector of light in the medium, the mutual interactions between the individual dipoles of the metasurface result in an electric and magnetic dipole moment $\boldsymbol{p}$ and $\boldsymbol{m}$ given by [32]:

$$\boldsymbol{p} = \alpha^{\mathrm{e}}\left[\boldsymbol{E_0} + \frac{k_0^2}{\varepsilon_0}\hat{G}^0 \cdot \boldsymbol{p}\right]$$

$$\boldsymbol{m} = \alpha^{\mathrm{m}}\left[\boldsymbol{H_0} + k_0^2\varepsilon_{\mathrm{d}}\hat{G}^0 \cdot \boldsymbol{m}\right].$$

Here, $\hat{G}^0 = \sum_{j=1}^{\infty} G_{0j}$ is the sum of the Green's functions of the single electric/magnetic dipoles at a position $j$ in the square lattice and takes into account effects arising from the periodic arrangement of the dipoles like the mutual interaction of the single electric and magnetic dipoles in the lattice and the appearance of diffraction. In the metamaterial limit, however, *i.e.* for lattice constants smaller than the wavelength of the incident light, the influence of radiative interactions between the dipoles can be neglected and the scattering is ultimately determined by the scattering properties of the individual particles [32] with the corresponding effective electric/magnetic polarizabilities $\alpha_{\mathrm{eff}}^{\mathrm{e,m}}$.

Starting from the electric and magnetic dipole moments given above, one can calculate the total electric far-field response outside of the silicon nanodisk metasurface which consist of the superposition of the incident field and the scattered field from the electric and the magnetic response of the nanodisk metasurface. For an incident polarization of the electric field vector along *x* or *y* direction, this finally results in a transmission coefficient of the silicon nanodisk metasurface given by:

$$t = 1 + \frac{ik_{\mathrm{d}}}{2A}(\alpha_{\mathrm{eff}}^{\mathrm{e}} + \alpha_{\mathrm{eff}}^{\mathrm{m}})$$

$A$ is the area of a unit cell, $k_{\mathrm{d}} = n_{\mathrm{d}} \cdot \omega/c_0$ is the wavevector of light in a medium with the refractive index $n_{\mathrm{d}}$, and $\alpha_{\mathrm{eff}}^{\mathrm{e}} = 1/(\varepsilon_0\varepsilon_{\mathrm{d}}/\alpha^{\mathrm{e}} - k_0^2\hat{G}^0)$ and $\alpha_{\mathrm{eff}}^{\mathrm{m}} = 1/(1/\alpha^{\mathrm{m}} - k_{\mathrm{d}}^2\hat{G}^0)$ are the effective (electric and magnetic) polarizabilities of the subwavelength nanodisk lattice.

To connect this expression with experimental transmittance spectra, we describe the dispersion of the effective electric and magnetic polarizabilities ($\alpha_{\mathrm{eff}}^{\mathrm{e}}$ and $\alpha_{\mathrm{eff}}^{\mathrm{m}}$) by Lorentzian line shapes given by:

$$\alpha_{\mathrm{eff}}^{\mathrm{e}} = \frac{\alpha_0^{\mathrm{e}}}{\omega_{\mathrm{e},0}^2 - \omega^2 - 2i\gamma_{\mathrm{e}}\omega}$$

$$\alpha_{\text{eff}}^{\text{m}} = \frac{\alpha_0^{\text{m}}}{\omega_{\text{m},0}^2 - \omega^2 - 2i\gamma_{\text{m}}\omega}$$

In a final step, the amplitudes of the effective polarizability amplitudes $\alpha_0^{\text{e,m}}$ have to be determined. Here, we use that, in the transmission spectrum, both the electric and the magnetic resonance are pronounced resonance dips at $\lambda = \lambda_{\text{el}}$ and $\lambda = \lambda_{\text{mag}}$ ideally reaching zero transmission when they are well separated from each other (see inset Fig. 3 a, for example). This is due to complete destructive interference between the incident wave and each of the two scattered fields in the ideal (loss-less) case and, therefore, allows us to determine the amplitudes $\alpha_0^{\text{e,m}}$ for the electric and magnetic scattering amplitudes by normalization of the complex scattering amplitudes $S^{\text{e,m}} = \frac{ik_d}{2A}\alpha_{\text{eff}}^{\text{e,m}}$. Finally we obtain $\alpha_0^{\text{e,m}} = \frac{4Ac_0}{n_d}\gamma_{\text{e,m}}$ and we can calculate an analytical expression for the (field-) transmission coefficient of an infinite subwavelength silicon-nanodisk lattice with electric and magnetic resonances:

$$t = 1 + \frac{2i\cdot\gamma_e\cdot\omega}{\omega_{e,0}^2 - \omega^2 - 2i\gamma_e\omega} + \frac{2i\cdot\gamma_m\cdot\omega}{\omega_{m,0}^2 - \omega^2 - 2i\gamma_m\omega}$$

The (field-) reflection coefficient can be obtained accordingly:

$$r = \frac{2i\cdot\gamma_e\cdot\omega}{\omega_{e,0}^2 - \omega^2 - 2i\gamma_e\omega} - \frac{2i\cdot\gamma_m\cdot\omega}{\omega_{m,0}^2 - \omega^2 - 2i\gamma_m\omega}$$

**Numerical calculations**

Numerical calculations are performed using the finite-integral frequency-domain (FIFD) solver provided by CST Microwave Studio. The silicon refractive index is taken as 3.5, that of the buried oxide layer as 1.45. We use unit cell boundary conditions and normal incidence plane-wave excitation. Referencing to a bare etched wafer structure is performed in Fig. 3 d-f in order to mimic the experimental measurement procedure.